\documentclass[a4paper,11pt]{article}
\usepackage{times}
\usepackage{amsmath,amssymb,amsfonts,amsthm}
\usepackage{graphicx}
\usepackage[colorlinks=true,urlcolor=blue]{hyperref}
\def\be{\begin{equation}}
\def\ee{\end{equation}}
\def\bea{\begin{eqnarray}}
\def\eea{\end{eqnarray}}
\def\lb{\label}
\def\ct{\cite}
\def\bi{\bibitem}
\def\gam{\gamma}
\def\eps{\epsilon}
\def\th{\theta}

\def\sig{\sigma}
\def\om{\omega}

\def\apj{{\em Astrophys. J.\/} }

\def\cqg{{\em Class. Quantum Grav.\/} }
\def\grg{{\em Gen. Rel. Grav.\/} }

\def\jmp{{\em J. Math. Phys.\/} }
\def\mn{{\em Mon. Not. R. Astron. Soc.\/} }
\def\prd{{\em Phys. Rev.\/} D }

\begin{document}
\title{\sc On the Newtonian Limit of the Weyl Tensor}
\author{{\sc J\"urgen Ehlers}${}^{1}$$\;$and
{\sc Thomas Buchert}${}^{2,3}$\thanks{E--mail: {\it
      buchert@obs.univ--lyon1.fr}} \\
{\small\em ${}^{1}$Max--Planck--Institut f\"ur Gravitationsphysik,
Albert--Einstein--Institut, Schlaatzweg 1, D--14473 Potsdam,
Germany (1996)} \\
{\small\em ${}^{2}$Ludwig--Maximilians--Universit\"at,
Theoretische Physik, Theresienstr.\ 37, D--80333 M\"unchen,
Germany (1996)}\\
{\small\em ${}^{3}$Universit\'e Lyon 1, Centre de Recherche Astrophysique de Lyon, CNRS UMR 5574} \\
{\small\em 9 avenue Charles Andr\'e, F--69230 Saint--Genis--Laval, France (current address)}}

\date{\normalsize{Publication of a Preprint: August 15, 1996}}
\maketitle
\begin{abstract}
In this note we wish to complement some recent work in the
cosmological literature concerning the Weyl conformal curvature tensor
and its parts. In particular, we shall give a clear--cut definition
of the Newtonian limits of electric and magnetic parts of the Weyl tensor.
We also discuss that only a subset of the relativistic
equations is needed to obtain a closed system of equations in the
Newtonian limit.

\end{abstract}
\begin{flushleft}
\end{flushleft}

\section{Introduction}
\lb{sec:intro}
In a recent paper Bertschinger \& Jain (1994) \ct{berjai1994}
attempted to derive a closed Newtonian system for the evolution of
fluid quantities (density, expansion, shear, etc.) by looking at
the Newtonian limit of the corresponding equations in general
relativity. The idea was to obtain ``local'' evolution equations
in terms of a coupled system of ordinary differential equations
(while only the initial data are constructed non--locally) in
order to simplify the solution of the problem of gravitational
motion and to answer the question whether the gravitational
collapse in general ends in sheet--like (oblate), or filamentary
(prolate) objects.

Thereafter, a considerable amount of work has been spent on
supporting or disproving this goal (Matarrese {\em et al.\/} 1994
\ct{matetal1994}, Bertschinger \& Hamilton 1994 \ct{berham1994},
Kofman \& Pogosyan 1995 \ct{kofpog1995}, Bruni {\em et al.\/}
1995 \ct{bruetal1995}, Lesame {\em et al.\/} 1996 \ct{lesetal1995},
Ellis \& Dunsby 1997 \ct{elldun1996}, 
Matarrese 1996 \ct{mat1996}, Matarrese \& Terranova
1996 \ct{matter1996}), and others. Some of these papers are even concerned
with new post--Newtonian theories (Bertschinger \& Hamilton 1994
\ct{berham1994}, see the discussion of that paper by Ellis \&
Dunsby 1997 \ct{elldun1996}).

We here wish to complement these works by focusing on two aspects
of the problem: firstly, we present a clear--cut derivation of
the Newtonian limit of fluid evolution equations in a
4--dimensional ``frame theory'' developed by one of us (Ehlers
1981, 1991 \ct{ehl1981,ehl1991}). This theory covers both
Einstein's and Newton's theory of gravitationally interacting
matter and allows to properly {\em define\/} the Newtonian limit.
We shall re--address questions related to the magnetic part of
the Weyl tensor.

Secondly, we discuss that only a subset of the general relativistic
equations is needed to obtain a closed Newtonian system, and we shall
establish the relation to the Newtonian Lagrangian framework
formulated first by Buchert \& G\"otz (1987) \ct{bucgoe1987}, in
Newtonian cosmology by Buchert (1989) \ct{buc1989}, and reviewed
by Buchert (1996) \ct{buc1996} and Ehlers \& Buchert (1997).

(Throughout this paper greek indices run through $0 \ldots 3$, and
latin indices through $1 \ldots 3$, the signature of the metric
is $(-,+,+,+)$.)

\section{The Newtonian limit of the Weyl tensor and its parts}
\lb{sec:weyl}
In the recent cosmological literature there have been discussions
on the role of the Weyl conformal curvature tensor, particularly
its ``magnetic'' part, in the Newtonian limit of general
relativity. These considerations suffer from the fact that claims
are made about that limit without any reference to, or use of, a
{\em definition\/} of that limit. However, a clear--cut (and
useful!) definition has been given long ago (K\"unzle 1976
\ct{kue1976}, K\"unzle \& Nester 1984 \ct{kuenes1984}, Ehlers
1981, 1991 \ct{ehl1981,ehl1991}, Brauer {\em et al.\/} 1994
\ct{braetal1994}).

Here we shall employ a ``frame--theory'' which covers, in a
common 4--dimensional spacetime formalism, Newton's as well as
Einstein's theory. In it, a parameter $\lambda=c^{-2}$ serves
to distinguish between the two theories, and the limit is taken
as $\lambda \to 0$. In that formalism one uses a temporal metric
$t_{\alpha\beta}$ and (inverse) spatial metric $s^{\alpha\beta}$,
related by:
\be
\lb{metric}
t_{\alpha\beta}s^{\beta\gam}=-\lambda\delta_{\alpha}^{\gam} \ .
\ee
In the GR--case, $\lambda > 0$ and
$g^{\alpha\beta}=s^{\alpha\beta}$,
$g_{\alpha\beta}=-\lambda^{-1}t_{\alpha\beta}$,
whereas in Newton's theory, $\lambda=0$ and $t_{\alpha\beta}
=t_{,\alpha}t_{,\beta}$, with $t$ the absolute time. In the
limit $\lambda \to 0$, the Lorentz metrics $g_{\alpha\beta}$,
$-\lambda g_{\alpha\beta}$ degenerate, corresponding to the fact
that the light cones ``open up'' and convert into the Newtonian
hyperplanes of constant absolute time (for details see: Ehlers
1981, 1991 \ct{ehl1981,ehl1991}).

In the notation of the frame--theory, the definition of the
Weyl tensor for $\lambda > 0$ reads:
\be
\lb{weyl}
C^{\alpha}{}_{\beta\gam\delta}
= R^{\alpha}{}_{\beta\gam\delta}
-\delta^{\alpha}_{[\gam}R_{\delta]\beta}
-\lambda^{-1}\left\{t_{\beta[\gam}R_{\delta]\varepsilon}
s^{\varepsilon\alpha}+\frac{1}{3}\,\delta^{\alpha}_{[\gam}
t_{\delta]\beta}R_{\lambda\mu}s^{\lambda\mu}\right\} \ .
\ee
This expression is meaningless for $\lambda=0$. However, if we
use the field equation
\be
R_{\alpha\beta}=8\pi G\left(t_{\alpha\gam}t_{\beta\delta}
-\frac{1}{2}\,t_{\alpha\beta}t_{\gam\delta}\right)T^{\gam\delta}
-\Lambda t_{\alpha\beta}
\ee
of the frame--theory, valid for $\lambda \geq 0$, to eliminate
$R_{\alpha\beta}$ from equation~(\ref{weyl}), we get:
\be
\lb{weyl2}
C^{\alpha}{}_{\beta\gam\delta}
= R^{\alpha}{}_{\beta\gam\delta}
-8\pi G\left\{\delta^{\alpha}_{[\gam}t_{\delta]\lambda}
t_{\beta\mu}T^{\lambda\mu}-t_{\beta[\gam}t_{\delta]\varepsilon}
T^{\varepsilon\alpha}-\frac{2}{3}\,\delta^{\alpha}_{[\gam}
t_{\delta]\beta}t_{\lambda\mu}T^{\lambda\mu}\right\} \ .
\ee
This formula is meaningful even for $\lambda=0$ (the
$\lambda^{-1}$--terms in (\ref{weyl}) cancel because of the
$\lambda$ in (\ref{metric})).
We therefore {\em define\/} the
Weyl tensor in the frame--theory by Equation~(\ref{weyl2}). This
definition is appropriate; for if a sequence of GR--solutions
has a Newtonian solution as a limit, then the limit of
$C^{\alpha}{}_{\beta\gam\delta}$ is indeed given by
Equation~(\ref{weyl2}).

In the Newtonian case, $\lambda=0$, (\ref{weyl2}) simplifies,
because of $t_{\alpha\beta}=t_{,\alpha}t_{,\beta}$, to
\be
\lb{weylnewton}
C^{\alpha}{}_{\beta\gam\delta}
= R^{\alpha}{}_{\beta\gam\delta}
-\frac{8\pi G}{3}\,\rho t_{,\beta}\delta^{\alpha}_{[\gam}
t_{,\delta]} \ .
\ee
The ``electric'' and ``magnetic'' parts of
$C^{\alpha}{}_{\beta\gam\delta}$ with respect to any 4--velocity
$u^{\alpha}$ follow from (\ref{weylnewton}); they read:
\bea
\lb{elenew1}
E^{\alpha}{}_{\gam}
& = & R^{\alpha}{}_{\beta\gam\delta}u^{\beta}u^{\delta}
-\frac{4\pi G}{3}\,\rho(\delta^{\alpha}_{\gam}
-u^{\alpha}t_{,\gam}) \ ; \\
\lb{magnew1}
H_{\alpha\gam}
& = & \frac{1}{2}\,\eta_{\alpha\beta\lambda\mu}s^{\mu\nu}
C^{\lambda}{}_{\nu\gam\delta}u^{\beta}u^{\delta}
= 0 \ .
\eea
The vanishing of $H_{\alpha\beta}$ in the Newtonian limit can be
understood more physically as follows: In GR, $H_{\alpha\beta}$
measures the relative rotation of nearby, freely falling
gyroscopes due to gravitomagnetism (Sachs 1960 \ct{sac1960},
Appendix\footnote{the relevant formula contains a misprint: the
correct version is $H^{a}{}_{[g]}=\frac{1}{2}\,\eta^{ai\ell m}
R_{\ell mjk}u_{i}\delta x^{j}u^{k}$.}). This effect is absent in
Newton's theory in which the parallelism of spatial vectors is
path independent, in other words, parallel gyroscopes remain
parallel if subject to nothing but inertia and gravity.

Investigation of the general case of exact solutions of general relativity with
$H_{\alpha\beta}=0$ and $\om_{\alpha\beta}=0$ has been attempted 
by Barnes \& Rowlingson (1989) \ct{barnesrow1989}. However, they have not
investigated the propagation of the dynamical constraint $H_{ab}=0$, which is needed
for statements about classes of exact solutions\footnote{This issue was considered later 
with the additional assumption $p=0 \Rightarrow
\dot{u}^{a}=0$ in van Elst {\em et al.\/} 1997 \ct{henk2}, where
for the general case without any symmetries and Weyl tensor of Petrov type I no complete results could be obtained.}.
Further investigations of such classes of motion termed ``silent
universes'' may be found in Croudace {\em et al.\/} 1994
\ct{croetal1994}, Bruni {\em et al.\/} 1995 \ct{bruetal1995},
Matarrese 1996 \ct{mat1996} (and ref. therein), and van Elst \&
Uggla 1997 \ct{hveugg1996}.

\section{Discussion of the equations in the Newtonian limit}
\lb{sec:newlim}
In 3--dimensional notation, (\ref{elenew1}) yields, as expected,
the Newtonian tidal tensor defined as the trace--free part of the
gravitational field tensor ($g_{i,j}$) (a comma denotes derivative with respect to Eulerian coordinates):
\be
\lb{elenew2}
E_{ij}:=g_{i,j}-\frac{1}{3}\,\delta_{ij}g_{\ell,\ell} \ ,
\ee
with
\be
E_{[ji]}=0 \ ; \qquad
E_{ii}=0 \ .
\ee
Like any Eulerian field, the Newtonian tidal tensor of the
gravitational field strength $\vec{g}$ can be written explicitly in
terms of Lagrangian coordinates as follows:
\be
\lb{elenew3}
E_{ij}=g_{i|k}J_{kj}^{-1}
-\frac{1}{3}\,\delta_{ij}g_{\ell|k}J_{k\ell}^{-1} \ ,
\ee
where a vertical slash denotes derivative with respect to
Lagrangian coordinates.

Introducing the diffeomorphic mapping $\vec{f}_{t}:
\vec{x}=\vec{f}(\vec{X},t)$, which sends fluid elements from
their (initial) Lagrangian position $\vec{X}$ to a point $\vec{x}$
in Eulerian space at time $t$, and using the expression for the
Jacobian of the inverse mapping $\vec{h}=\vec{f}^{-1}$,
\bea
& & \vec{g}[\vec{x},t]:=\ddot{\vec{f}}(\vec{h}[\vec{x},t],t) \ ,
\qquad
J:=\det(f_{i|k}) \ , \\
& & h_{j|\ell}
=\frac{1}{2J}\,\eps_{\ell pq}\eps_{jrs}f_{p|r}f_{q|s} \ ,
\eea
we can write the tidal tensor explicitly in terms of $\vec{f}$:
\be
\lb{elenew4}
E_{ij}=\frac{1}{2J}\left(\eps_{jpq}{\cal J}(\ddot{f}_{i},f_{p},f_{q})
-\frac{1}{3}\,\eps_{opq}{\cal J}(\ddot{f}_{o},f_{p},f_{q})
\delta_{ij}\right) \ .
\ee
Thus, any trajectory field $\vec{f}$ which obeys the
Lagrange--Newton system (Buchert \& G\"otz 1987 \ct{bucgoe1987}
($\Lambda=0$) and Buchert 1989 \ct{buc1989} ($\Lambda \neq 0$);
Ehlers \& Buchert 1997 \ct{ehlbuc1996}),
\bea
\lb{detid1}
{\cal J}(\ddot{f}_{j},f_{j},f_{k}) & = & 0 \ , \\
\lb{detid2}
{\cal J}(\ddot{f}_{1},f_{2},f_{3})+{\cal J}(\ddot{f}_{2},f_{3},f_{1})
+{\cal J}(\ddot{f}_{3},f_{1},f_{2})-\Lambda J
& = & -4\pi G\stackrel{0}{\rho} \ ,
\eea
determines the evolution of the tidal tensor via
(\ref{elenew3}).

In (\ref{detid1}) and (\ref{detid2}), ${\cal J}({\cal A},{\cal B},
{\cal C})$ denotes the functional determinant of any three
functions ${\cal A}(\vec{X},t)$, ${\cal B}(\vec{X},t)$ and
${\cal C}(\vec{X},t)$ with respect to Lagrangian coordinates
$\vec{X}$, $\stackrel{0}{\rho}$ is the initial density field and
$\Lambda$ the cosmological constant.

We may use Equation~(\ref{elenew4}) to state the Lagrange--Newton
system~(\ref{detid1}) and (\ref{detid2}) in a different way: it is
equivalent to the condition that $E_{ij}$ is symmetric and
tracefree (Buchert 1996 \ct{buc1996}): We insert (\ref{detid2})
into (\ref{elenew4}) and write
\be
\lb{elenew5}
E_{ij}=\frac{1}{2J}\,\eps_{jpq}{\cal J}(\ddot{f}_{i},f_{p},f_{q})
-\frac{1}{3}\left(\Lambda-\frac{4\pi G}{J}\,\stackrel{0}{\rho}
\right)\delta_{ij} \ .
\ee
Then,
\bea
E_{[ij]}=0 & \Leftrightarrow & (\ref{detid1})\ ; \\
E_{ii}=0 & \Leftrightarrow & (\ref{detid2}) \ .
\eea
This demonstrates that no equation involving $H_{ij}$ is needed
in the Newtonian limit to close the system. Indeed, for
$H_{\alpha\beta}=0$, the relativistic system of equations for the fluid
variables expansion ($\th$), vorticity ($\om_{\alpha\beta}$),
shear ($\sig_{\alpha\beta}$), ``electric'' ($E_{\alpha\beta}$)
and ``magnetic'' ($H_{\alpha\beta}$) parts of the Weyl tensor
--- derived by one of us (Ehlers 1961; translated 1993
\ct{ehl1961}) and Tr\"umper 1965 \ct{tru1965}, and reviewed
by Ellis 1971 \ct{ell1971} --- reduces to the system
\bea
\lb{rhodot}
\dot{\rho} & = & -\th\rho \ , \\
\dot{\th} & = & \Lambda-\frac{1}{3}\th^{2}+2(\om^{2}-\sig^{2})
-4\pi G\rho \ , \\
\lb{omdot}
\dot{\om}^{\alpha} & = & -\frac{2}{3}\th\om^{\alpha}
+ \sig^{\alpha}{}_{\beta}\om^{\beta} \ , \\
\lb{sigdot}
\dot{\sig}_{\alpha\beta} & = & -\sig_{\alpha\gam}\sig^{\gam}{}_{\beta}
-\om_{\alpha}\om_{\beta}+\frac{1}{3}h_{\alpha\beta}(2\sig^{2}-\om^{2})
-\frac{2}{3}\th\sig_{\alpha\beta}-E_{\alpha\beta} \ , \\
\dot{E}_{\alpha\beta} & = & -h_{\alpha\beta}\sig^{\gam\delta}
E_{\gam\delta}-\th E_{\alpha\beta}+E_{\gam(\alpha}
\om_{\beta)}{}^{\gam}+3E_{\gam(\alpha}\sig_{\beta)}{}^{\gam}
-4\pi G\rho\sig_{\alpha\beta} \ ,
\eea
where $h^{\alpha\beta}=g^{\alpha\beta}+u^{\alpha}u^{\beta}$ is the
spatial projection tensor. Recall that a solution of this closed set of equations
delivers at best an approximation that needs to be controlled. Resulting solutions
will have to be subjected to the propagation constraint 
$\dot{H}_{\alpha\beta}=0$ to find the exact solution classes. This constraint will likely leave only
highly symmetric solutions like the Lema\^\i tre--Tolman--Bondi and the Szekeres models, since,
in general, a non--vanishing magnetic part of the Weyl tensor is generated by evolving a system
with initially vanishing magnetic part. 

In the Newtonian limit $\lambda \to 0$ the spatial parts of the
equations (\ref{rhodot})--(\ref{sigdot}) are equivalent to those
derived from Newtonian theory (compare Szekeres \& Rankin 1977
\ct{szeran1977} and the discussions by Kofman \& Pogosyan 1995
\ct{kofpog1995}, Lesame {\em et al.\/} 1996 \ct{lesetal1995},
Buchert 1996 \ct{buc1996} and Matarrese 1996 \ct{mat1996}).
The Lagrange--Newton--System (\ref{detid1}) and (\ref{detid2})
can be already obtained from (\ref{rhodot})--(\ref{omdot})
together with the decomposition $\sig_{ij}+\om_{ij}
+\frac{1}{3}\th\delta_{ij}=v_{i,j}$.

\bigskip
\noindent
{\it Note added 2009:} Some examples of Newtonian limits of relativistic spacetimes have been published by Ehlers 1997 together with more details on the Newtonian limit in the frame--theory formalism used here \ct{ehlexamples1997}.
Furthermore, it is possible to write the Einstein equations in terms of symmetry conditions imposed on the electric and magnetic parts of the Weyl tensor similar to writing the Lagrange--Newton system  (\ref{detid1}, \ref{detid2}),
since both parts of the Weyl tensor contain all information. The investigation of this formulation together with the corresponding Newtonian limit will be published elsewhere. Related remarks on the Newtonian limit in the framework of Cartan's formulation of Einstein's equations may be found in Buchert (2008) \ct{buchert:de}, Sect. 4.2.1.

\section*{\small Acknowledgments}
In 1996, TB was supported by the ``Sonderforschungsbereich 375--95 f\"ur
Astro--Teilchenphysik der Deu\-tschen Forschungsgemeinschaft''. He
would like to thank the Albert--Einstein--Institut in Potsdam for
generous hospitality during a working visit in 1996, where this work has
been written. TB would also like to thank Henk van Elst for digging out the preprint, as well as for help and comments during the 
preparation of this manuscript.

\addcontentsline{toc}{section}{References}

\end{document}